\documentclass[twocolumn,preprintnumbers,amsmath,amssymb,floatfix,superscriptaddress,nofootinbib]{revtex4}

\usepackage{graphicx}
\usepackage{dcolumn}
\usepackage{bm}
\usepackage{hyperref}
\usepackage{color}
\usepackage{fontawesome} 
\usepackage{comment}
\usepackage{slashed}
\usepackage{tikz}
\usetikzlibrary{positioning,decorations.pathmorphing}
\usepackage{braket}
\usepackage{xspace}

\definecolor{deepblue}{rgb}{0.2,0.2,0.8}
\definecolor{deepred}{rgb}{0.8,0.2,0.2}
\definecolor{deeporange}{rgb}{0.8,0.5,0.2}
\definecolor{deepgreen}{rgb}{0.2,0.6,0.2}

\newcommand{\be}{\begin{equation}}
\newcommand{\ee}{\end{equation}}

\definecolor{linkcolor}{rgb}{0.7752941176470588, 0.22078431372549023, 0.2262745098039215}

\begin{document}

\title{Dark wounds on icy moons: Ganymede's subsurface ocean as a dark matter detector}

\author{William DeRocco}
\email{derocco@umd.edu}
\affiliation{Maryland Center for Fundamental Physics, University of Maryland, College Park, 4296 Stadium Drive, College Park, MD 20742, USA}
\affiliation{Department of Physics \& Astronomy, The Johns Hopkins University, 3400 N. Charles Street, Baltimore, MD 21218, USA}

\date{\today}

\begin{abstract}
Dark matter in the form of macroscopic composites is largely unconstrained at masses of $\sim 10^{11}- 10^{17}$ g. In this mass range, dark matter may collide with planetary bodies, depositing an immense amount of energy and leaving dramatic surface features that remain detectable on geological timescales. In this paper, we show that Ganymede, the largest Jovian moon, provides a prime target to search for dark matter impacts due to its differentiated composition and Gyr-old surface. We study the effects of dark matter collisions with Ganymede first with analytic estimates, finding that in a large region of parameter space, dark matter punches through Ganymede's conductive ice sheet, liberating sub-surface material. This sub-surface material may be compositionally different from the surface ice, providing a key observable with which to discriminate asteroid impacts from those caused by dark matter. We confirm our analytic estimates with dedicated simulations of dark matter impacts using iSALE, a multi-material impact code. We then discuss potential detection prospects with two missions currently en route to the Jovian system, Europa Clipper and JUICE, finding that these missions may have the ability not only to identify signs of life on the Galilean moons, but signs of dark matter as well.
\end{abstract}

\maketitle

\section{Introduction}
\label{sec:intro}

The nature of dark matter (DM) constitutes one of the largest open problems in physics. While the majority of attention has focused on the search for new particles with intrinsically weak interactions, an alternative paradigm, known as \textit{macroscopic dark matter}, suggests that dark matter may instead be composed of high-mass macroscopic objects with sufficiently low number density to evade current detection efforts \cite{Jacobs2014}. Examples of such dark matter candidates include primordial black holes \cite{pbh,Hawking:1971ei,Carr:1974nx,Chapline:1975ojl,Carr:2020gox,Green:2020jor}, solitons \cite{Lee:1986tr,Lee:1991ax,Kawana:2021tde,Hong:2020est,Hardy:2014mqa,Wise:2014jva,Wise:2014ola,Gresham:2017cvl,Chang:2018bgx,Flores:2020drq,Flores:2023zpf,DelGrosso:2023trq,Lu:2024xnb,Coleman:1985ki,Kusenko:1997ad,Kusenko:1997si,Kusenko:2001vu,Kasuya:1999wu,Frieman:1988ut}, strangelets \cite{Witten:1984rs,Madsen:1986jg,Madsen:1998uh,Bai:2018dxf,Bai:2018vik,zhitnitsky_nonbaryonic_2003}, and more. Several model-agnostic searches have been performed for these candidates, largely ruling out DM structures with geometric cross sections with the Standard Model at masses below 100 g and above $10^{22}$ g \cite{Sidhu2020}. However, a large open parameter space remains for such DM candidates in the range of $10^{12} -10^{22}$ g and densities ranging from atomic (1 g/cm$^3$) to nuclear ($10^{14}$ g/cm$^3$). This is an exceptionally challenging parameter space to probe, as the high mass leads to a correspondingly small number density in the Galaxy, with, e.g., only one $10^{14}$ g DM constituent passing  through the Earth every $\approx 10^{5}$ years. As a result, any attempt to search for such DM requires enormous detectors with long integration times.

Using astrophysical bodies such as stars and planets as DM detectors provides a means of overcoming this challenge. Due to their long lifetimes and large volumes, stars and planets can record the history of their interactions with DM, enabling a new means of probing this parameter space. This insight has been used to constrain macroscopic DM at high densities through white dwarfs \cite{Graham2015}, neutron stars \cite{Sidhu2020}, and red giants \cite{Dessert2022}, however these targets suffer from decreased sensitivity towards low densities. Alternatives, such as the use of Solar System bodies \cite{Yalinewich:2021fdr,Picker2025}, have been suggested to fill this gap, but at present suffer from considerable theoretical uncertainty and a poor ability to discriminate from conventional Standard Model processes.

In this paper, we will show that Ganymede, the largest Galilean moon, has the potential to probe this interesting region of parameter space. With a size larger than Mercury and a surface that has largely remained the same for over two billion years, it presents a compelling target for the search for DM. Critically, due to its compositionally-differentiated subsurface layers, DM collisions with the moon can release deep subsurface material that traditional impacts cannot, providing a critical signature by which to discriminate DM interactions from conventional ones. We demonstrate this first with analytical arguments based on simple scaling laws before verifying these predictions with dedicated multi-material impact simulations. Our results show that future missions, such as the European Space Agency's Jupiter Icy Moons Explorer (JUICE) \cite{Grasset2013}, may be sensitive to macroscopic DM over a large parameter space.

The paper is organized as follows. In Sec. \ref{sec:ganymede}, we introduce the Jovian moons and describe upcoming missions to characterize their surfaces and composition. In Sec. \ref{sec:impacts}, we describe the qualitative behavior of dark matter impacts on Ganymede's surface and provide analytic estimates for relevant length and time scales. In Sec. \ref{sec:simulations}, we describe a set of dedicated multi-material simulations of these impacts, then show the results of these simulations in Sec. \ref{sec:results}. We discuss the implications of these results in Sec. \ref{sec:discussion} before concluding in Sec. \ref{sec:conc}.

\section{Ganymede}
\label{sec:ganymede}

Icy moons are a class of natural satellites found around several of the gas giants in the Solar System. They are characterized by surfaces composed primarily of ice, though their interiors are thought to contain deep sub-surface oceans of liquid water. As such, they are an exciting locale in which to search for life in the Solar System, and several current and future missions are set to characterize these worlds. 

In particular, two space missions will explore the Galilean moons of Jupiter over the coming decade. These are the NASA-led Europa Clipper mission \cite{EuropaClipper} and the ESA-led Jupiter Icy Moons Explorer (JUICE) mission \cite{Grasset2013}. As the name suggests, Europa Clipper will largely focus on Europa, one of the smaller Galilean moons. Surface features on Europa hint at a tidally-heated liquid water ocean beneath its icy crust, making it a prime target in the search for life in the Solar System.

Though Europa may be the prime target in the search for life, it is not the ideal target in the search for dark matter. Its small size and active resurfacing limit its ability to record traces of dark matter impacts. Ganymede, its larger cousin in the Jovian system, provides a more compelling target. Ganymede is the largest of the Galilean moons, with a radius of 2631 km. It is believed to be composed primarily of silicate rock and water in  differentiated layers. Current models suggest a conductive ice Ih crust 12 -- 14 km deep \cite{Bjonnes2022} that transitions into a convective ice layer that extends to $\approx 100$ km below the surface. Beneath this is a subsurface ocean of a thickness $\approx 250$ km \cite{Grasset2013}, then, potentially, another layer of exotic, high-pressure ice VI. This sits atop a rocky mantle, beneath which lies an iron core.

The surface is dominated by two different types of terrain. The darker regions that cover roughly one-third of Ganymede's surface have been dated to be on the order of 4.2 billion years old \cite{Neukum1998}, while the brighter regions are thought to be the product of a resurfacing event in the more recent past, the exact mechanism of which is unknown. These younger regions, which constitute roughly two-thirds of Ganymede's surface, are believed to be on the order of 2 billion years old \cite{Zahnle2003}.

The first mission to characterize the composition of Ganymede's surface was NASA's Galileo spacecraft \cite{Carlson1992}. Through infrared reflectance imaging by the spacecraft's Near-Infrared Mapping Spectrometer (NIMS), the mission found evidence for non-ice materials on the surface such as a variety of hydrated mineral salts that were suggested to have originated from the upwelling of briny layers in the subsurface ocean at some point in the past \cite{McCord2001}. More recent ground-based observations have confirmed these signatures and provide evidence that chlorinated salts from deep layers of the ice sheet and liquid ocean may be the dominant chemical species on the surface \cite{King2022}. Furthermore, laboratory studies of sodium chloride hydrates suggest that hyperhydrated salts (e.g. 2NaCl$\cdot$17H$_2$O) could form in the subsurface ices and be transported near to the surface through convective processes. The identification of hyperhydrated chlorides on the surface from their infrared reflectance spectra may provide a signpost for areas in which material recently upwelled from deeper layers of the ice \cite{Journaux2023}. NASA's Juno mission has also recently provided high resolution infrared spectra of Ganymede's surfaces down to scales of less than a kilometer per pixel, revealing local variations in Ganymede's surface composition that further highlight regions of possible ocean-to-surface transport in the past \cite{Tosi2024}.

JUICE's MAJIS will take high-precision infrared spectra of the entire surface of Ganymede at resolutions ranging from 3 km/px globally to less than 100 m/px for targets of interest \cite{Tosi2024Review}. This will be augmented by the JANUS camera, which will be capable of taking optical observations in multiple bands at a resolution of 10 m/px. JANUS high-resolution and stereoscopic imaging of morphological structures will allow for chemical abundances observed by MAJIS to be correlated with particular surface features \cite{Tosi2024Review}. The subsurface composition of such features will be further characterized by RIME, JUICE's ice-penetrating radar payload, which will probe the ice sheet down to a depth of nine kilometers at a resolution of a few meters. These observations will look for compositional changes with depth and structures related to melt generation and injection during impact cratering \cite{Heggy2017}. This large suite of complementary experiments will provide unprecedented insight into the composition and morphology of small-scale structures on Ganymede's surface.

\begin{table*}[t]
    \centering
    \begin{tabular}{c|c|c|c}
        Parameter & Value & Description & Ref. \\
        \hline
        \hline
        $\rho_\text{ice}$ & 0.92 g/cm$^3$ & Density of ice & \cite{Bjonnes2022}\\
        $g$ & 1.43 m/s$^2$ & Surface gravity of Ganymede & \cite{Bjonnes2022}\\
        $R_\text{Ganymede}$ & 2631 km & Radius of Ganymede & \cite{Bjonnes2022} \\
        $P_\text{vap}$ & 70 GPa & Shock vaporizaton pressure & \cite{Kraus2011} \\ 
        $P_\text{melt}$ & 3.5 GPa & Shock melt pressure & \cite{Kraus2011} \\ 
        $Y$ & 110 MPa & Strength parameter at high pressure & \cite{Bray2009} \\
         $z_\text{ice}$ & 12 km & Conductive ice depth & \cite{Bjonnes2022} \\
        $\tau$ & 2 Gyr & Ganymede resurfacing time & \cite{Zahnle2003}\\
        $v_\text{DM}$ & 270 km/s & Velocity of dark matter & \cite{Catena2012} \\

    \end{tabular}
    \caption{Parameter values used for the analytic impact estimates in Sec. \ref{sec:impacts}.}
    \label{tab:parameters}
\end{table*}

\begin{figure*}
    \includegraphics[width=\textwidth]{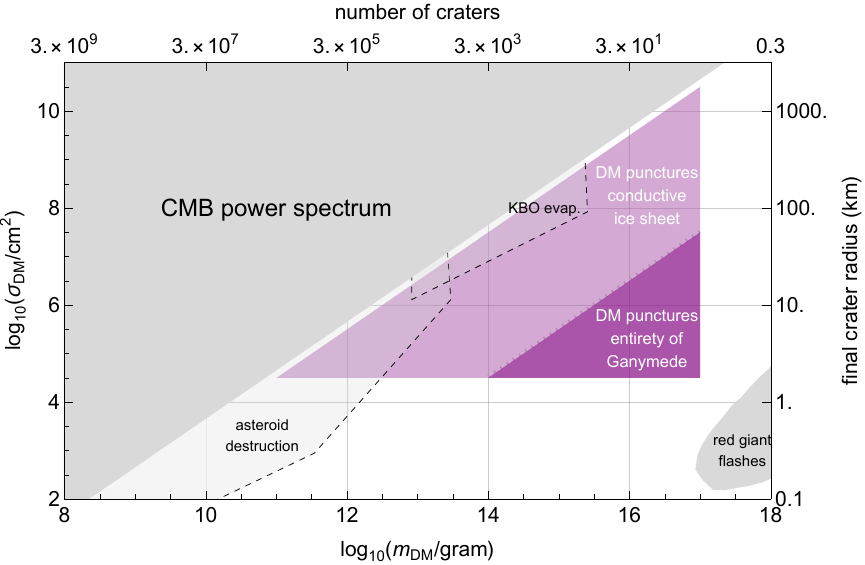}
    \caption{The region in which dark matter impacts may leave detectable surface features on Ganymede. Gray regions correspond to existing constraints \cite{Sidhu2020,Picker2025,Dessert2022}. The light purple region is the region in which the dark matter has sufficient initial energy to penetrate through the 12 km conductive ice layer on Ganymede's surface. The darker purple region corresponds to when the dark matter penetrates the entirety of Ganymede, leaving both an entrance and exit wound. Rough estimates for the approximate number of craters and crater radius from Eqs. \ref{eq:nimpacts} and \ref{eq:cratersize} are provided on the top and right axes respectively, rounded to one significant figure for readability. JUICE's global spectral mapping will enable the detection of features down to kilometer-scale over the majority of Ganymede's surface \cite{Tosi2024Review}, providing the resolution and coverage necessary to identify craters of interest throughout the purple-shaded parameter space.}
    \label{fig:moneyplot}
\end{figure*}

Ganymede's large radius and long resurfacing time allow it to record signatures of impacts with dark matter over geologic time, as pointed out recently by \cite{Santarelli:2025eye}. Using Ganymede's radius and surface age, we can estimate the expected number of such impacts that it has undergone during a resurfacing time. As displayed in Table \ref{tab:parameters}, we adopt $v_\text{DM} = 270$ km/s, $R_\text{Ganymede} = 2631$ km, and $\tau = 2$ Gyr. With this, we find that $N_\text{impacts} \approx \frac{\rho_\text{DM}}{m_\text{DM}} \pi R_\text{Ganymede}^2 v_\text{DM} \tau$, which yields 
\begin{equation}
\label{eq:nimpacts}
    N_\text{impacts} \approx 2.6 \left( \frac{m_\text{DM}}{10^{17}\,\text{g}}\right)^{-1}.
\end{equation}
As such, our parameter space is limited to macroscopic dark matter composites below about $10^{17}$ g (Fig. \ref{fig:moneyplot}). At lower masses, however, dark matter impacts may have left a large number of wounds on the surface of Ganymede with distinct morphology and composition in comparison to standard asteroid impacts. In the following section, we will explore these features analytically before comparing our results to simulation in Sec. \ref{sec:simulations}.

\section{Dark matter impacts}
\label{sec:impacts}

There is an extensive body of literature on impact cratering, with seminal work deriving scaling laws for crater size and structure performed dating back decades \cite{Melosh1989,Holsapple1993}. However, almost all existing work to date has focused on the resulting cratering from point-like depositions of energy near the impact surface. This is a reasonable approximation in the context of traditional asteroid impacts, as upon impact, the weak structural integrity of typical asteroids causes them to rapidly disintegrate, releasing all of their energy as a small, effectively point-like detonation near to the target surface.

The picture differs significantly for macroscopic dark matter impacts. In order to evade existing cosmological constraints, the density of macroscopic dark matter must be much greater than that of typical Solar System bodies ($\mathcal{O}(1)$ g/cm$^3$), with densities as high as $\sim 10^{8}$ g/cm$^3$ in regions of interest for this paper. Though we remain agnostic to the microphysical model, this density implies that generically, whatever force binds the dark matter to keep it at such extreme densities is much stronger than the typical forces binding rock and ice. As a result, the dark matter impactor would not to disintegrate upon collision with a planet or moon, and would instead punch a deep borehole, hollowing out a cylindrical cavity in its wake. In some extreme cases, the impactor may even fully penetrate a planetary target, carving a wound all the way through the planet before exiting the other side.

Furthermore, typical asteroid collisions rarely occur at velocities much exceeding $\sim \mathcal{O}(10)$ km/s. Dark matter impactors, however, are expected to enter the Solar System at roughly the Galactic virial velocity (270 km/s) \cite{Catena2012}, which is orders of magnitude larger than typical collision speeds. As such, the typical kinetic energy of an incoming dark matter impactor can be three to four orders of magnitude larger than an asteroid of the equivalent mass.

Though the dynamics of dark matter impacts differ considerably from traditional asteroid impacts, we expect that the phases of impact cratering are similar to those of traditional impacts \cite{Melosh2011}. As such, we break the process of dark matter impacts into three stages: (1) energy deposition and shock propagation, (2) expansion and excavation flow, and (3) collapse and late-time material transport. The remainder of this section will be devoted to an analytic treatment of each of these phases, along with quantitative estimates for impacts with Ganymede's surface using the fiducial parameter values found in Tab. \ref{tab:parameters}.

\subsection{Energy deposition and shock propagation}

Since the DM velocity is much larger than typical shock speeds in ice and rock, we treat the energy deposition arising from a DM passage as an effective line-source explosion with constant energy deposited per unit length of 
\begin{equation}
\label{eq:dedx}
    dE/dx \approx \sigma \rho_\text{ice} v_\text{DM}^2\\ = 2.1\times10^{14}~\text{J/m}\left(\frac{r_\text{DM}}{1\,\text{m}}\right)^2,
\end{equation}
where $\sigma = \pi r_\text{DM}^2$ is the cross-sectional area of an impactor with radius $r_\text{DM}$, $\rho_\text{ice} = 0.92$ g/cm$^3$ is the density of ice near the surface, and $v_\text{DM} = 270$ km/s is the impact velocity, which we take to be constant throughout the duration. 
Given this energy deposition rate, we can estimate the parameter space in which the dark matter impactor passes fully through Ganymede. The penetration depth $D$ is roughly given by $D \approx E_0/(dE/dX)$, where we take an initial kinetic energy is $\frac{1}{2} m_\text{DM} v_\text{DM}^2$ and we use $3$ g/cm$^3$ as a typical average density encountered by dark matter passing through the entirety of Ganymede. This yields a penetration depth of 
\begin{equation}
\label{eq:penetrationdepth}
    D \approx 5300\,\text{km}\left(\frac{m_\text{DM}}{10^{14}\,\text{g}}\right) \left(\frac{r_\text{DM}}{1\,\text{m}}\right)^{-2},
\end{equation}
which is slightly larger than Ganymede's diameter ($\approx5262$ km). Hence, we see that dark matter impactors with $m_\text{DM}/r_\text{DM}^2 \gtrsim 3\times 10^{13} \, \text{g/m}^2$ (or $\approx 2.4\times10^{7}$ g/cm$^3$ assuming the macro is a sphere) may fully puncture Ganymede and exit the other side (Fig. \ref{fig:moneyplot}, dark purple region). Similarly, we find that all DM above $m_\text{DM}/r_\text{DM}^2 \gtrsim 3\times 10^{10} \, \text{g/m}^2$ penetrates the $\approx 12 - 14$ km thick conductive ice shell and could potentially bring convective ice with unique compositional signatures to the surface (Fig. \ref{fig:moneyplot}, light purple region). Note that it is by coincidence that that this limit very nearly coincides with the existing constraints from the CMB power spectrum shown in Fig. \ref{fig:moneyplot}. 

The passage of the dark matter immediately produces a cylindrical region of radius a few $r_\text{DM}$ at immense pressures, on the order of $\rho_\text{ice} v^2$. This triggers a shock wave that propagates outward through the surrounding ice at a velocity well exceeding the sound speed in ice. The pressure drops rapidly both due to energy losses and due to the geometric scaling with increasing radius. The scaling behavior for a shock wave produced by a cylindrical line charge can be estimated largely through dimensional analysis, yielding
\begin{equation}
\label{eq:vshock}
  v_\text{shock}(r) \propto \left(\frac{dE/dx}{\rho_\text{ice}}\right)^{1/2} r^{-1}
\end{equation}
and
\begin{equation}
\label{eq:Pshock}
   P_\text{shock}(r) \sim \rho_\text{ice} v_\text{shock}(r)^2 \sim \left(dE/dx\right) r^{-2},
\end{equation}
which matches the scaling of more sophisticated analytic models \cite{Lin1954}. Substituting fiducial values for Ganymede, we have
\begin{equation}
   P_\text{shock}(r) \approx 6.7\times10^4~\text{GPa} \left(\frac{r_\text{DM}}{r}\right)^2.
\end{equation}

This pressure is well above the shock vaporization pressure and melt pressure in ice. We adopt the values for the shock pressure at which ice is completely vaporized (melted), $P_\text{vap}$ ($P_\text{melt}$) given in Ref. \cite{Kraus2011}, taking $P_\text{vap} = 70$ GPa and $P_\text{melt} = 3.5$ GPa. Given these and Eq. \ref{eq:Pshock}, we can solve for the radius surrounding the DM trajectory at which ice is vaporized/melted. This yields:
\begin{equation}
   \frac{R_\text{vap}}{r_\text{DM}} \sim v_\text{DM}\sqrt{\frac{\rho_\text{ice}}{P_\text{vap}}} \approx 31 
\end{equation}
and
\begin{equation}
   \frac{R_\text{melt}}{r_\text{DM}} \sim v_\text{DM}\sqrt{\frac{\rho_\text{ice}}{P_\text{melt}}} \approx 138.
\end{equation}
These results should be taken as highly approximative, but they suggest that the immediate consequence of a dark matter impact with Ganymede would be to vaporize (melt) a column of ice $\mathcal{O}(10)$ ($\mathcal{O}(100)$) times the impactor radius, setting the stage for the following phase of cratering.

\subsection{Expansion and excavation flow}

Though some energy is used to melt and vaporize ice in the regions nearest to the impactor's trajectory, the remaining energy goes to setting up a shock wave that quickly races away from the column of impact. As suggested by Eq. \ref{eq:Pshock}, the peak pressure is expected to fall roughly quadratically with distance from impact column, a scaling that we verify with simulations in Sec. \ref{sec:simulations}. As the shock propagates outward and the pressure decreases, its thermodynamic irreversibility induces a net outward excavation flow at an order-one fraction of the shock velocity \cite{Melosh2011}. The outward excavation flow can be stopped by either hydrostatic pressure or by the target material's strength. In the case of Ganymede, the structural integrity of the ice halts the excavation flow in the upper $\sim80$ km of the icy crust before transitioning to a hydrostatic-dominated regime beneath. Since we are primarily interested in the behavior in the upper $\approx20$ km of the ice, we estimate the maximum width of the transient cavity by finding the radius at which the pressure from the excavation flow is of order $Y$, the strength parameter for ice, which we take to be $Y = 110$ MPa \cite{Bray2009}. Setting the pressure from equation \ref{eq:Pshock} equal to this value, we have
\begin{equation}
\label{eq:rmaxice}
    \frac{R_\text{max, ice}}{r_\text{DM}} \sim \sqrt{\frac{\rho_\text{ice} v_\text{DM}^2}{Y}} = 780. 
\end{equation}
This phase of impact ends when the transient cavity (in our case, a borehole) has attained its maximum radius and begins to collapse, which occurs on  a timescale of roughly
\begin{equation}
    \Delta t_\text{expansion} \sim\frac{\rho_\text{ice} v_\text{DM} r_\text{DM}}{Y} = 2.26\, \text{s} \left(\frac{r_\text{DM}}{1\,\text{m}}\right).
\end{equation}

\subsection{Collapse and late-time material transport}

After the transient cavity reaches its maximum radius, it begins to collapse as melted and brecciated ice falls inwards. 
The collapse occurs in the deepest regions of the borehole first, owing to the higher hydrostatic pressures and narrower borehole. As the lower parts collapse back inwards, they create a pressure gradient that drives material upwards, forming a phenomenon known as a \textit{Worthington jet} \cite{Worthington1897,Worthington1900}. Worthington jets have been studied extensively in the field of elastoviscous hydrodynamics and analytic models have been developed to characterize their behavior \cite{Gekle2010}. The most salient result from these studies is that for a wide variety of initial conditions, the resulting jet typically has a width of $\mathcal{O}(0.1) \times r_b$, where $r_b$ is roughly the point at which the inward radial flow from the collapsing column is converted to upwards axial flow. This point moves outward as the jet evolves, but is typically between $0.1 - 0.4$ times the width of the transient cavity. In the context of DM impacting Ganymede, this provides a crude means of estimating the volume of material contained within this jet that is ultimately transported to the surface. We are interested primarily in the material that is transported from beneath the conductive ice sheet (deeper than 12 km), hence we only wish to estimate the amount of material that the jet has accumulated prior to reaching this depth. We therefore roughly estimate the volume to be $\approx \mathcal{O}(0.01) \pi r_\text{max,ice}^2 l_\text{jet}$, where $l_\text{jet} = z_\text{pinch} - z_\text{ice}$ depends on the the point at which the jet begins to form due to ``pinching off'' from hydrostatic instabilities ($z_\text{pinch}$). This quantity is difficult to estimate without simulations, but will likely lie somewhere between $15 - 80$ km, with the lower value being when the ice becomes largely convective and the higher being when hydrostatic forces begin to dominate material strength, narrowing the transient cavity. This yields
\begin{equation}
M_{<12\,\text{km}} \approx 1.2\times10^9 \, \text{kg} \left(\frac{r_\text{DM}}{1\,\text{m}}\right)^2 \left(\frac{z_\text{pinch}}{20\,\text{km}}\right).
\end{equation}
However, we wish to caution that this estimate may considerably overestimate the true jet mass, as it neglects the effects of the early vaporization on removing material that otherwise would have contributed to the jet.

The timescale over which the system relaxes and the base of the jet is brought to the surface is of order the gravitational free-fall timescale to the pinch-off point, $\Delta t_\text{jet} \approx 2\sqrt{z_\text{pinch}/g}$. This follows from the fact that the collapse of the transient borehole proceeds due to hydrostatic pressure, driving an upward flow of order $\frac{dz}{dt}(z) \approx \sqrt{g z}$ \cite{Gekle2010}. Inverting this expression and integrating from $z = z_\text{pinch}$ to $z=0$ produces the same result. For Ganymede, this yields something on the order of
\begin{equation}
    \label{eq:tpinch}
    \Delta t_\text{jet} \approx 4\,\text{min} \left(\frac{z_\text{pinch}}{20 \, \text{km}}\right)^{1/2}.
\end{equation}

Given the large uncertainties in the approximations made above, it is this phase of cratering that most requires simulation in order to make robust quantitative estimates. As such, in the following Section, we verify the qualitative features of dark matter impacts described in the previous sub-sections and make more robust quantitative estimates through the use of simulations.

\section{Simulations}
\label{sec:simulations}

In order to have a more accurate quantitative understanding of potential surface features on Ganymede, we use the multi-material shock physics code iSALE \cite{SALE,Collins2004,Wunneman2006}. We adopt the same input parameters and material properties for Ganymede's ice sheet as displayed in Table 1 of Ref. \cite{Bjonnes2022}, which ran comprehensive simulations of traditional impact cratering on Ganymede to estimate the depth of its conductive crust. We adopt a convective ice temperature of 260 K and thermal gradient of 10 K/km, which the authors found best reproduced the observed distribution of crater morphologies. Our simulation region consists of an $840\times2000$-cell grid of high resolution $10\,\text{m} \times 10\,\text{m}$ cells and an extension region with extension factor 1.06 that consists of 134 cells above the surface and 402 cells beyond the high-resolution region in the radial direction. All high-resolution cells contain one tracer particle, which is used to evaluate the volume and mass of displaced material. The simulation files are all available on Zenodo \cite{ZenodoCite}.

\begin{figure}
    \includegraphics[width=\linewidth]{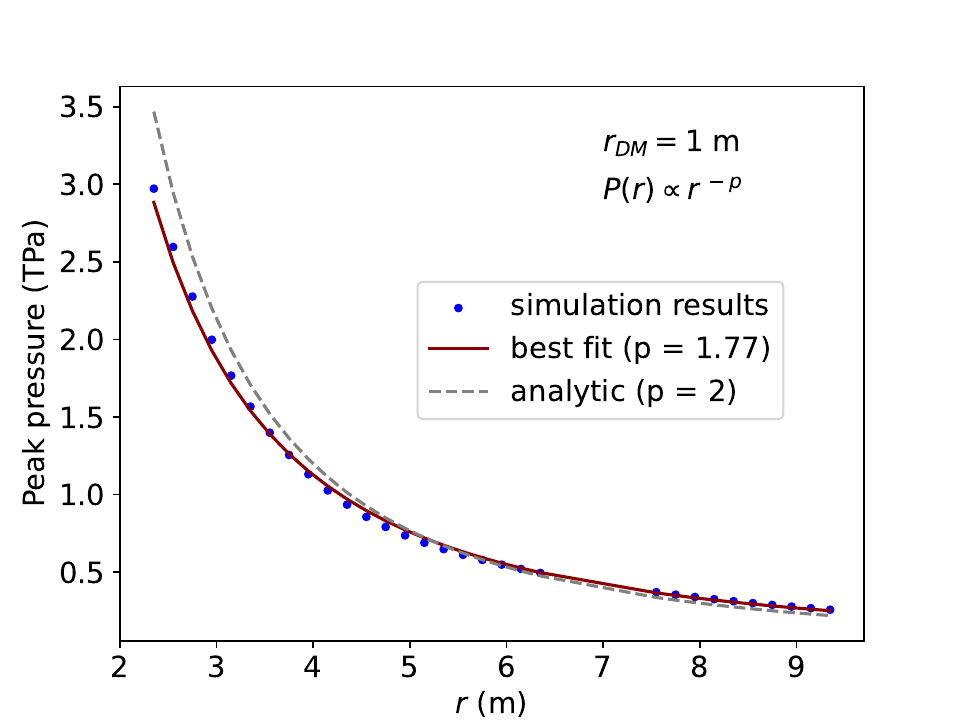}
    \caption{Peak pressure as a function of radius from impactor column. Blue dots show the measured values for a simulation of a 1-meter radius dark matter impactor impinging on Ganymede's surface. The red solid curve shows a best-fit power-law to the data while the gray dotted curve shows the prediction from the analytic scaling arguments of Sec. \ref{sec:impacts}, indicating good correspondence.}
    \label{fig:peakpressures}
\end{figure}

Unlike simulations of traditional impact cratering by asteroids, the enormous pressures and energies involved with dark matter impacts, coupled with their small spatial size and high velocity, introduce a large hierarchy of scales that prevents direct simulation of the impactor itself. Instead, as described in Sec. \ref{sec:impacts}, we choose to model the impact as a uniform linear energy deposition. At the start of the simulation, we manually deposit an energy density per unit mass of $(dE/dx)/(\pi \rho_\text{ice} (r_\text{DM}/10)^2) = (v_\text{DM}/10)^2 = 7.2\times10^6$ J/kg in a cylinder of radius $10 r_\text{DM}$ that extends to the bottom of the simulation region. Note that we have chosen to distribute the initial impact energy over a column ten times the width of the dark matter in order reduce the hierarchy of scales involved and make these simulations computationally tractable. This assumes that the peak pressure should fall off as $P \propto r^{-2}$, as suggested by Eq. \ref{eq:Pshock}, which we confirm by running a dedicated short-duration simulation in which the dark matter energy deposition and radius are set to their true values ($r = 1$ m, specific energy = $7.2 \times 10^{10}$ J/kg). Due to the large hierarchy of scales, these simulations are extremely costly and take a long time, however the results confirm that the analytic prediction of a  $P \propto r^{-2}$ scaling is roughly correct, with our data suggesting a best-fit scaling of $P \propto r^{-1.77}$, as shown in Fig. \ref{fig:peakpressures}.  If anything, this means that our assumption for the initial energy deposition may slightly underestimate the size of the transient cavity, but a wider cavity would presumably only increase the amount of material that could be brought to Ganymede's surface.

\begin{figure*}
    \centering
    \includegraphics[width=\textwidth]{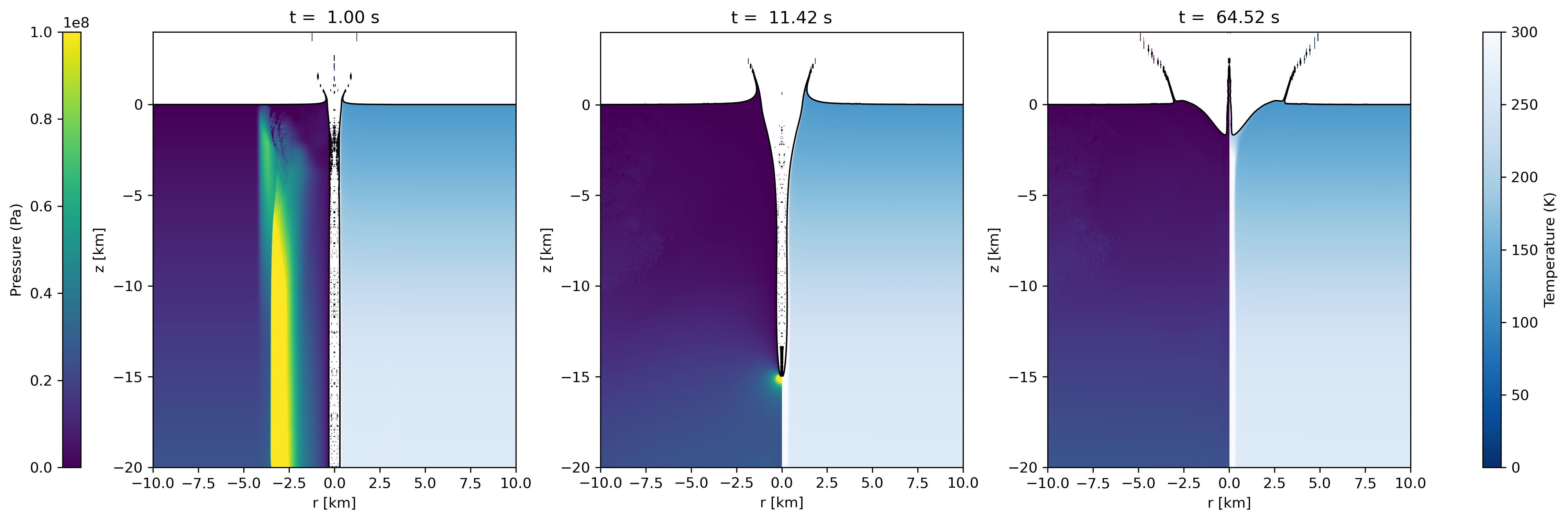}
    \caption{Primary results from simulations. The three panels show the evolution of a 1-meter radius dark matter impactor colliding with Ganymede at three times. At $t = 1$ s (leftmost panel), the shock has formed and propagated $\approx 3$ km, leaving behind a transient borehole of radius $\approx 0.3$ km. At $t = 11.42$ s (middle panel), the borehole has begun to collapse from the bottom up, forming a jet of subsurface material. At $t=64.52$ s (rightmost panel), this jet reaches the surface and escapes, potentially leading to observable compositional differences in the final relaxed crater.}
    \label{fig:difftimes}
\end{figure*}

\begin{figure}[hbtp]
    \includegraphics[width=\linewidth]{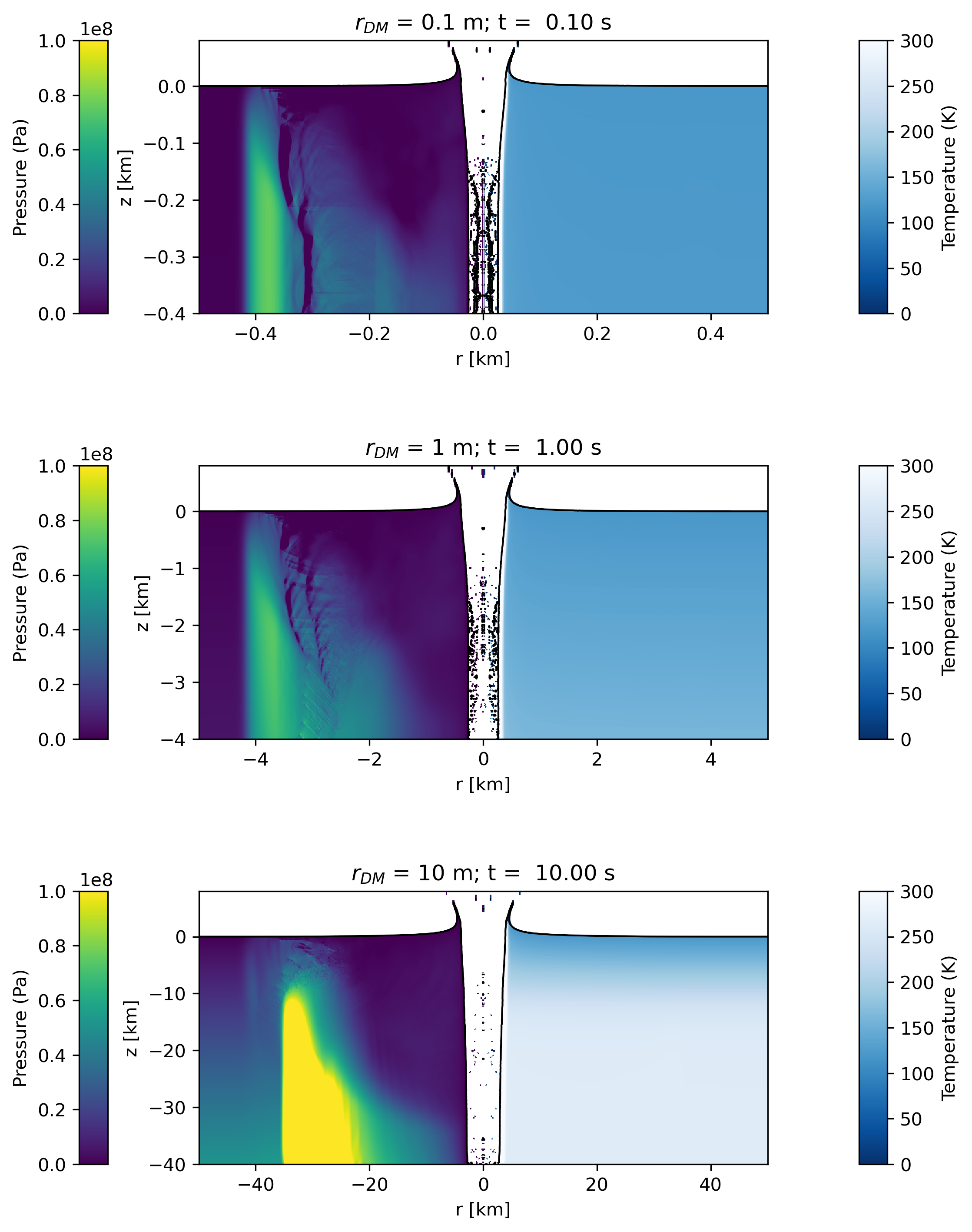}
    \caption{Simulation results for impactors of varying size at corresponding times post-impact. The results show that both the transient cavity radius and timescale for formation scale approximately linearly with the radius of the impactor. Note the changing scale on the axes between the panels.}
    \label{fig:diffsizes}
\end{figure}

\section{Results}
\label{sec:results}

We present the main results from the simulations in Fig. \ref{fig:difftimes}. These panels show the pressure and temperature profiles of Ganymede's ice shell at three different times following the impact of a 1-meter radius impactor. The first panel (left) is taken at $t=1$ s, and shows the initial formation of the transient borehole and the propagation of the shock outwards into the ice. The cavity reaches a maximum radius of $\approx 0.3$ km, which is slightly smaller than the analytic estimate (Eq. \ref{eq:rmaxice}), but of the correct order of magnitude. The second panel corresponds to 11.42 seconds after impact, and we see the partial collapse of the borehole and the initial formation of the jet (material just above the pressure spike at the point of collapse). The final panel corresponds to 64.52 seconds after impact. By this time, the jet has propagated to the surface, bringing melted deep subsurface ice with it. This is slightly faster than predicted from hydrostatic collapse (Eq. \ref{eq:tpinch}), but is again of the right order of magnitude. At this point, the crater radius has reached approximately 3 km at the surface. This should scale linearly with the impactor radius (see discussion at the end of this Section), hence we find
\begin{equation}
\label{eq:cratersize}
    D \approx 6 \,\text{km} \left(\frac{r_\text{DM}}{1 \,\text{m}}\right)
\end{equation}
where $D$ is the final crater diameter. This may slightly underestimate the final crater size, as this both neglects the contribution from the ejecta blanket and from the final stages of surface relaxation, but it provides a conservative estimate of the scale.

Due to computational cost, our simulation was terminated at this point. However, by this point, the behavior of interest has already been captured and we are able to estimate the amount of material brought to the surface during collapse.
We define subsurface upwelling in our simulations as any material that originated below 12 km beneath the surface and is brought to within 2 km of the surface (the depth of the transient crater at the end of the simulations, see Fig. \ref{fig:difftimes}) by the end of the simulation. We ignore all material that is initially vaporized and escapes, instead focusing on late-time entrained material. Under this definition, we find that $\approx 5 \times 10^7$ kg ($\approx$ 50,000 cubic meters) of subsurface ice is brought to the surface and escapes. This may be an underestimate of the total volume that would be liberated in a realistic impact due to the finite size of our simulation region, however we anticipate that the liberated volume is fairly independent of penetration depth, as the majority of the material originates from just below the start of the convective ice layer. It escapes at the surface with an average velocity of 222 m/s, which is well below the escape velocity from the surface of Ganymede (2.741 km/s), hence the material will jet outwards before ultimately settling back down in the vicinity of crater.

Under the assumption that, by geometric arguments, the liberated volume should scale roughly with with the cross-sectional area of the transient cavity, we find an approximate relation
\begin{equation}
M_{<12\, \text{ km}}  \approx 5\times10^7 \, \text{kg} \left(\frac{r_\text{DM}}{1 \, \text{m}}\right)^2
\end{equation}
for the amount of material brought up from below the transition to the convective ice layer. This is significantly less than the estimate made in Sec. \ref{sec:impacts}, possibly due to the loss of material by vaporization, however the result still shows that DM has the potential to liberate a large volume of fluid from deeper layers of Ganymede's ice.

Though our primary simulation mimics the effects of a 1-meter radius impactor impinging on Ganymede at a velocity of 270 km/s, these results can be rescaled to other impactor sizes through the equations presented in Sec. \ref{sec:impacts}. To confirm this, we ran simulations for both 0.1 meter impactors and 10 m impactors as well, finding that features such as the transient crater diameter scale as expected from analytics. We display the results of these simulations side-by-side in Fig. \ref{fig:diffsizes}. Moving from top to bottom, we show a simulation of a 0.1 m impactor at 0.1 s after impact, a 1 m impactor at 1 s, and a 10 m impactor at 10 s. The three snapshots are nearly indistinguishable, up to a slight difference in the peak pressure for the 10 m impactor, providing strong evidence for the validity of rescaling our fiducial $r_\text{DM} = 1$ m results.

\section{Discussion}
\label{sec:discussion}

The results of the simulations show that it is possible for dark matter impactors at least as large as 1 meter in radius to bring subsurface material to Ganymede's surface. Given this, we display the parameter space in which crater composition and morphology on Ganymede's surface may provide signatures of dark matter impacts in purple in Fig. \ref{fig:moneyplot}. The horizontal axis corresponds to the mass of the impactor while the vertical axis is its cross-sectional area ($\sigma_\text{DM} = \pi r_\text{DM}^2$). We additionally provide approximate values for the final relaxed crater radius taken from simulations (number of craters on Ganymede's surface) along the right-hand (top) side of the plot. These should be taken largely as suggestive order of magnitude estimates intended to provide a sense of scale and frequency.

We cut off the parameter space above $10^{17}$ g, beyond which there would be no impacts over the 2 billion year exposure time (see Sec. \ref{sec:ganymede}), and restrict the impactor to be $> 1$ m in radius, since this is the lowest radius that was simulated for a sufficient duration to display subsurface jetting. Finally, the parameter space is restricted to impactors of sufficient density to penetrate the 12 km conductive ice layer, a bound that only by chance roughly aligns with existing constraints from the cosmic microwave background \cite{Sidhu2020}. We also display bounds from both other Solar System objects \cite{Picker2025} and from red giant flashes \cite{Dessert2022} in gray, however the former largely remain approximative and may not fully rule out the indicated parameter space.

As described in Sec. \ref{sec:ganymede}, JUICE is expected to perform global spectral mapping of Ganymede's surface with MAJIS at an average spatial resolution of $\approx3$ km/px, with the ability to resolve features an order of magnitude below this in regions of interest \cite{Tosi2024Review}. This is a 1 to 2 order of magnitude improvement on current global spectral maps obtained by Galileo's NIMS, which were limited to resolutions of $\approx 100$ km/px \cite{Tosi2024Review}. As such, JUICE will provide the first opportunity to characterize the kilometer-scale variations in surface composition that could be used as indicators of dark matter impacts and will be able to identify such indicators over the entirety of the purple-shaded region in Fig. \ref{fig:moneyplot}.

It is important to keep in mind that above a crater radius of $\mathcal{O}(10)$ km, traditional asteroid impacts may also puncture the conductive crust \cite{Bjonnes2022}, hence distinguishing a dark matter impact from a traditional asteroid impact becomes more challenging. There may, however, still be significant compositional differences if the dark matter penetrates into deeper layers, bringing different kinds of subsurface material to the surface, and ground-penetrating radar such as that used by JUICE may help discriminate a roughly hemispherical melt volume from a deep, narrow, cylindrical melt column. Note that craters several hundreds of kilometers in diameter have been observed on Ganymede, hence a lack of craters at an apparent size does not place a strong constraint on this dark matter parameter space.

Perhaps one of the most interesting features would be the presence of exit wounds in the parameter space in which dark matter fully penetrates Ganymede and exits the other side (dark purple in Fig. \ref{fig:moneyplot}). Given the fact that the energy deposition is effectively instantaneous, modeling it as a linear charge deposition still applies, hence we expect the exit wound to largely mimic the late-time features of the entry wound. In extreme cases, it may even be possible to correlate these roughly antipodal surface features, which would provide very strong evidence for a dark matter origin.

\section{Conclusion}
\label{sec:conc}

In this paper, we have shown that for a large range of parameter space, dark matter impacts on Ganymede may be able to bring material from deep subsurface layers up to the surface. If this material possesses different compositional elements to that of the surface ice (e.g. hyperhydrated salts, see Sec. \ref{sec:ganymede}), such features could be identified by high-resolution spectral imagers aboard  missions such as JUICE and Europa Clipper that are currently en route to the Jovian system. Of particular note would be small craters ($\lesssim 10$ km) with an anomalously large amount of melt volume displaying composition significantly different to that of surrounding regions, since such features would be difficult to explain via conventional cratering. Follow-up of such craters with high-resolution imaging and subsurface radar like that carried aboard JUICE may reveal them to be signposts of past dark matter impacts.

Due to the 2D nature of iSALE simulations, we have restricted our attention to vertical impacts, however a more realistic treatment would include the effects of impactors at various impact angles. Shallow impacts could potentially leave an even richer phenomenology of surface features, e.g. an ejecta blanket oriented predominately on one side of the crater with a pool of subsurface material on the opposite side. These features may provide new means of discriminating dark matter impact sites from traditional craters, though we leave a full exploration of this to future work.

We acknowledge that the results presented in this paper are speculative. For dark matter impacts to provide discernible signatures requires not only for subsurface material to be brought to the surface, but for Ganymede's deeper layers to have significant compositional differences. The inner structure of Ganymede is largely unknown; characterizing it is one of the primary motivators behind missions such as Europa Clipper and JUICE. So although the work presented here is speculative at present, in the next few years, our knowledge of the icy moons in our Solar System will improve considerably. As our understanding of them evolves, so will our understanding of their potential as dark matter detectors. And perhaps, in coming years, we may find that the missions current en route to the Jovian moons reveal not only new hints to the origin of life, but new hints to the origin of the Universe as well.


\medskip

\acknowledgments{This work has been several years in the making and has evolved considerably over that time. WD wishes to extend an immense thanks to Sam Cabot, whose talk on hypervelocity impacts on the lunar surface started this entire project and who provided extensive help on iSALE simulations in the early stages of this work. WD further wishes to thank the myriad colleagues he has had the pleasure of discussing this with, with particular thanks to the group at TUM for mountaintop discussions early on.  Furthermore, WD would like to thank Marc Neveu for his insight on melt tubes on Europa and  Lynnae Quick for suggesting Ganymede as a target. The author also gratefully acknowledges the developers of iSALE-2D, including Gareth Collins, Kai Wünnemann, Dirk Elbeshausen, Tom Davison, Boris Ivanov, and Jay Melosh. This work was supported by NSF grant PHY-2210361 and the Maryland Center for Fundamental Physics.} 

\bibliography{refs}

\end{document}